\documentclass[a4paper]{jpconf}
\usepackage{graphicx}

\begin{document}
\title{\Large \bf The electron electric dipole moment enhancement factors of Rubidium and Caesium atoms}

\author{H S Nataraj$^{1}$, B K Sahoo$^{2}$, ~B P Das$^{1}$, ~R K Chaudhuri$^{1}$ ~and~ D Mukherjee$^{3}$ \\}

\address{$^1$ Indian Institute of Astrophysics, Bangalore 560 034, India} 
\address{$^2$ Max Planck Institute for the Physics of Complex Systems, 01187 Dresden, Germany} 
\address{$^3$ Indian Association for the Cultivation of Sciences, Calcutta 700 032, India}

\ead{nataraj@iiap.res.in, bijaya@mpipks-dresden.mpg.de, das@iiap.res.in, rkchaudh@iiap.res.in and pcdm@mahendra.iacs.res.in}

\begin{abstract}
The enhancement factors of the electric dipole moment (EDM) of the ground states of two paramagnetic atoms; rubidium (Rb) and caesium (Cs) which are sensitive to the electron EDM are computed using the relativistic coupled-cluster theory and our results are compared with the available calculations and measurements. The possibility of improving the limit for the electron EDM using the results of our present work is pointed out.\\ 
\end{abstract}

%\renewcommand{\baselinestretch}{1.5}
%\normalsize

\section{Introduction}
With excellent progress in the experimental techniques and the computational advancements in the field of high precision atomic physics in recent years, it has now become possible to undertake rigorous studies to test the validity of many models of particle physics which question the very fundamental laws of physics \cite{ginges, maxim}. One such challenge for both theorists and experimentalists is the simultaneous existence of parity (P) and time reversal (T) violations giving rise to the electric dipole moment (EDM) of an elementary particle like the electron \cite{sandars, sandars2}.
It is interesting for theorists because it is a direct proof of the time reversal violation in nature and also it has profound consequences ranging from sub-atomic physics to cosmology \cite{kazarian, branco}. It will be a break through for experimentalists as it requires considerable ingenuity and the state of the art technology to measure the tiniest electromagnetic moment to date.\\

An atom can have EDM even in the absence of any electromagnetic field due to an intrinsic EDM of any of its constituent particles like electrons, nucleons, quarks or also due to P \& T violating electron-nucleus interactions. In this paper, we have restricted ourselves to the calculation of the atomic EDM arising from the electron EDM. The intrinsic EDM of an electron interacts with the internal electrostatic fields in an atom, thereby enhancing the atomic EDM. Following the ideas first discussed by Schiff \cite{schiff}, Sandars explicitly demonstrated that, even if the electron has a non-zero EDM, the EDM of an atom in the non-relativistic case will be zero \cite{sandars2}. However, he went on to show that, in the relativistic treatment of the problem the atomic EDM does not vanish \cite{sandars2,salpeter}. Incidentally, the paramagnetic atoms, namely those with one valence electron, are predominantly sensitive to the electron EDM \cite{ginges, maxim}. The electron EDM contribution to atomic EDM in paramagnetic atoms, is shown to be proportional to $\vec d_e\,\alpha^2 \,Z^3$ \,\cite{ginges, maxim}, where, $Z$ is the atomic number, $\alpha$ is the fine structure constant and $\vec d_e$ is the electron EDM given by, $\vec d_e = d_e\, \beta \,\vec \sigma$ where, $d_e$ is the magnitude of the electron EDM, $\beta$ is the Dirac matrix and $\vec \sigma$ is the spin of the electron. The atomic EDM points in the direction of total angular momentum of the atom. Thus, the atomic EDM due to the electron EDM would be significantly enhanced in heavy paramagnetic atoms. \\

Indeed, we calculate a dimensionless quantity called {\em enhancement factor} ($R$) which is the ratio of an atomic EDM ($D_a$) to electron EDM $(d_e)$ using a highly correlated all-order relativistic many-body theory known as the relativistic coupled-cluster (RCC) theory. In this paper, we present $R$ due to the electron EDM of the ground states of Rb and Cs atoms. The current best limit for the electron EDM comes from an experiment on atomic thallium (Tl) \cite{regan} and prior to that an experiment on atomic Cs had set the limit for the electron EDM \cite{sudha}. New experiments on the EDM of atomic Rb and Cs using the modern techniques such as laser cooling and trapping are underway in two different laboratories in the world \cite{weiss, heinzen}. In combination with the results of our enhancement factors reported in this paper, they could improve the limit of the electron EDM by at least two orders of magnitude.\\

In the following section, we have briefly described the RCC method used in calculating the enhancement factors of Rb and Cs atoms. In section 3, we have presented the results and compared them with the existing calculations. In section 4, we have drawn the conclusions based on our results and described their implications for the particle physics.\\

\section{RCC Method}
At first, we obtain a reference wave function ($|\Phi_0\rangle$) for the closed-shell configuration with $N-1$ electrons, $N$ is the total number of the electrons in the system including
a valence electron ($v$), by solving the Dirac-Hartree-Fock (DF) equations with the Coulomb interaction; the Hamiltonian is given by,
\begin{eqnarray}
H_0\, =\, \sum_i \{ c\, \alpha_i \cdot p_i + (\beta_i - 1)\,m\,c^2 + V_{nuc}\,(r_i)\} + \sum_{i < j}\,V_c\,(r_{ij})
\label{H0}
\end{eqnarray}
where, $c$ is the speed of light in vacuum, $\alpha$ and $\beta$ are the Dirac matrices, $V_{nuc}$ is the nuclear potential.\\

In the framework of the RCC theory, we construct the
exact wave function ($|\Psi_0 \rangle$) for the closed-shell configuration as, 
\begin{eqnarray}
|\Psi_0 \rangle = e^{T^{(0)}}\, |\Phi_0\rangle,
\label{eqn2}
\end{eqnarray}
where, $T^{(0)}$
%$T^{(0)}= T1^{(0)} + T2^{(0)} = sum_{a,p} a_p^\dagger a_a t_a^p + sum_{ab,pq}a_p^\dagger a_q^\dagger a_b a_a t_{ab}^{pq} $ 
are the excitation operators from the core electrons; the details of which are given in \cite{lindgren}. It is the sum of all
single, double, triple and multiple excitations of occupied electrons. For a
single valence open-shell atomic system, which is of our interest, we construct the new DF wave function $\vert \Phi_v \rangle$ by appending
the valence electron $v$ to the closed-shell DF wave function $|\Phi_0\rangle$ and obtain the
corresponding reference state; i.e. $|\Phi_v\rangle = a_v^{\dagger} |\Phi_0\rangle$ where, $a_v^{\dagger}$ is the creation operator. The exact wave function for the corresponding valence electron system can be expressed
as, 
\begin{eqnarray}
\vert \Psi_v\rangle = e^{T^{(0)}}\{e^{S_v^{(0)}}\} \vert\Phi_v\rangle,
\label{eqn3}
\end{eqnarray}
where, $S_v^{(0)}$ corresponds to the excitation operator for the valence and valence-core orbitals.
Since the systems considered in this paper have only one valence electron, the non-linear terms in $S_v^{(0)}$ will
not exist and the above wave function reduces to the form \cite{mukherjee},
\begin{eqnarray}
\vert \Psi_v\rangle = e^{T^{(0)}}\{1+S_v^{(0)}\} \vert\Phi_v\rangle .
\label{eqn4}
\end{eqnarray}

It is impractical to consider all the correlated excitations in our calculation because of the huge requirement of the computer memory. It has been found that the RCC theory with both single and double excitations is quite successful in incorporating the maximum correlation effects because of its all-order nature. Hence, we have used in this calculation, the CC method with only the single and the double excitations (CCSD method); i.e, $T^{(0)} = T1^{(0)} + T2^{(0)}$ and $S_v^{(0)} = S1_v^{(0)} + S2_v^{(0)}$. However, we have also considered the non-linear terms up to $(T1^{(0)})^4$. In our earlier works, we have given the details of the working equations to determine both the $T^{(0)}$ and $S_v^{(0)}$ amplitudes and the ionization potential for the corresponding valence electron $v$ \cite{mukherjee}. We have also considered the leading triple excitations 
using CCSD(T) approximation, the details of which can be found in \cite{kaldor}, to improve the accuracy of the results. \\

The unperturbed RCC operator amplitudes, $T^{(0)}$ and $S_v^{(0)}$ are solved in two steps: first, we solve for the closed-shell 
amplitudes, $T^{(0)}$ using the following equations;
\begin{eqnarray}
\langle \Phi_{0}\vert\, \overline{H_0^\mathcal{N}} \,\vert \Phi_{0}\rangle &=& \Delta E_{corr}  \\
\langle \Phi_{0}^* \vert\, \overline{H_0^\mathcal{N}}\, \vert \Phi_{0} \rangle &=& 0 ,
\label{eqn5}
\end{eqnarray}
where, $\Delta E_{corr} = E_g - E_{DF}$ is the correlation energy, which is the difference between ground state energy and DF energy, for the closed-shell system, $|\Phi_{0}^* \rangle$ represents all possible singly and doubly excited states
with respect to the reference state, $|\Phi_{0} \rangle$ and we define
$\overline{H_0^\mathcal{N}} = e^{-T^{(0)}}\, H_0^\mathcal{N}\, e^{T^{(0)}}$ where, $H_0^\mathcal{N}$ is the normal ordered atomic Hamiltonian given by  $H_0^\mathcal{N} = H_0 - \langle \Phi_0 \vert H_0 \vert \Phi_0 \rangle$. The above two equations are solved simultaneously to get $T^{(0)}$ amplitudes self-consistently.\\

Then, we solve the following two equations
to obtain $S_v^{(0)}$ amplitudes and the ionization potential (IP) of the valence electron; 
\begin{eqnarray}
\langle \Phi_v \vert\, \{\overline{H_0^\mathcal{N}}\}_{op} \{1+S_v^{(0)}\}\vert \Phi_v\rangle &=& -\Delta E_v  \\
\langle \Phi_v^{*} \vert\, \{\overline{H_0^\mathcal{N}}\}_{op} \{1+S_v^{(0)}\}\vert \Phi_v\rangle &=& -\langle \Phi_v^{*}\vert S_v^{(0)} \vert \Phi_v\rangle \,\Delta E_v ,
\label{eqn6}
\end{eqnarray}
where, $\Delta E_v$ is the negative of the IP of the valence electron $v$ and
$|\Phi_v^{*}\rangle$ represent all possible singly and doubly excited states
with respect to the state $|\Phi_v\rangle$. Here, $\{\overline{H_0^\mathcal{N}}\}_{op}$ represent all the
open contracted operators from $\overline{H_0^\mathcal{N}}$ with $N$ electrons. The
fully contracted terms are subtracted from the total energy $E_v$ to get
$\Delta E_v$.\\

In the absence of any external field, we have intrinsic electron EDM contribution to atomic EDM given by, $d_e \sum_i \beta\, \vec \sigma_i \cdot \vec E_{int}$ where, $E_{int} = - \nabla \cdot \{\sum_i V_{nuc}\,(r_i) +\, \sum_{i<j}V_c\,(r_{ij})\}$ is the total internal electric field in an atom.
The energy shift due to the EDM can only be measured in the presence of an externally applied electric field. The applied field also induces the EDM to an atom  which is given by, $\sum_i \{ e \vec r_i \,+\, d_e\, \beta\, \vec \sigma_i \} \cdot \vec E $, where, $\vec E$ is the applied electric field.\\

Thus, the total atomic Hamiltonian in the presence of EDM as a perturbation is given by,
\begin{eqnarray}
H\, =\, H_0\, + \, H_{EDM}
\label{totalH}
\end{eqnarray}
where, $H_0$ is the unperturbed Hamiltonian given by eqn. (\ref{H0}) and  
$H_{EDM}$ is the EDM perturbed Hamiltonian given by,
\begin{eqnarray}
H_{EDM}\, =\, -\,d_e \sum_i \beta\, \vec \sigma_i\,\cdot\, \vec E_{int}.
\end{eqnarray}

The EDM perturbed Hamiltonian will effectively reduces to the one-body operator given by,
\begin{equation}
H_{EDM}^{eff}\, = \, 2\, i\, c\, d_e \sum_i \gamma_5 \,\beta\, \vec {p}_i\,^2
\end{equation}
where, $\beta$ and $\gamma_5$ are the Dirac matrices, $\vec p_i$ is the four-momentum of the electron, $\hbar$ is the modified Planck's constant (i.e, $\frac{h}{2\pi}$). The $H_{EDM}$ mixes the atomic states of opposite parities but
 with the same angular momentum. As its strength is sufficiently weak, we
consider only up to the first-order perturbation in wave function and
the modified atomic wave function for the valence electron state '$v$' is given by,
\begin{equation}
|\Psi_v' \rangle = |\Psi_v^{(0)} \rangle + d_e\, |\Psi_v^{(1)} \rangle
\end{equation}

In the RCC ansatz, the cluster operators for calculating the perturbed wave function are given by
\begin{eqnarray}
T = T^{(0)} + d_e\, T^{(1)} \nonumber \\
S_v =  S_v^{(0)} + d_e \,S_v^{(1)}
\end{eqnarray}
where, $T^{(1)}$ and $S_v^{(1)}$ are the first order corrections to the unperturbed cluster operators  $T^{(0)}$ and $S_v^{(0)}$, respectively.\\ 

The perturbed cluster amplitudes, $T^{(1)}$ and $S_v^{(1)}$ are obtained by solving the following equations respectively in a self consistent manner;
\begin{eqnarray}
\langle \Phi^* |\,\overline{H_0^{\mathcal{N}}}\, T^{(1)} + \overline{H_{EDM}^{eff}}\,| \Phi_0 \rangle &=& 0. \nonumber 
\end{eqnarray}
\begin{eqnarray}
\langle\Phi_v^*|(\overline{H_0^{\mathcal{N}}}- \Delta E_v)\, S_v^{(1)} + (\overline{H_0^{\mathcal{N}}}\, T^{(1)} + \overline{H_{EDM}^{eff}})\{1+S_v^{(0)}\}|\Phi_v\rangle=0 \nonumber
\end{eqnarray}

The expectation value of an atomic EDM for the state $|\Psi_v' \rangle$ (with the normalized unperturbed wave function) is given by,
\begin{eqnarray}
D_a \cong \frac {\langle \Psi_v'|\, D\, |\Psi_v' \rangle } {\langle \Psi_v^{(0)}|\Psi_v^{(0)} \rangle} 
\end{eqnarray}
where, $D = e z$ is the electric dipole operator. On substituting the expressions of the wave functions and keeping the terms only up to linear order in perturbed operators, we get the electron EDM enhancement factor ($R = \frac{D_a}{d_e}$) as,
\begin{eqnarray}
R = \frac{\langle \Phi_v |\left\{ \{1+ S_v^{(0)^\dagger}\} \overline{D^{(0)}} \{ T^{(1)} + T^{(1)} S_v^{(0)} + S_v^{(1)}\} + \{ S_v^{(1)^{\dagger}} + S_v^{(0)^{\dagger}} T^{(1)^{\dagger}} + T^{(1)^{\dagger}}\} \overline{D^{(0)}} \{1+ S_v^{(0)}\} \right\} |\Phi_v \rangle }{\langle \Phi_v |\, e^{T^{(0)^\dagger}}\, e^{T^{(0)}} \,+\, S_v^{(0)^{\dagger}}\, e^{T^{(0)^{\dagger}}}\, e^{T^{(0)}}S_v^{(0)}\, | \Phi_v \rangle}
\end{eqnarray}

where, $\overline{D^{(0)}} = e^{-T^{(0)}}\, D\, e^{T^{(0)}}$.

\section{Results \& discussion}

We have computed the single particle energies and wave functions self consistently using the Dirac-Fock (DF) method \cite{rajat} and calculated the EDM enhancement factors as explained above for the ground states of Rb and Cs atoms. The Table-1 summarizes these results. The DF result is based on the independent particle model (IPM) approach where the electron correlation effects are neglected. We have included all possible single, double and partial triple excitations of all-order using a highly correlated many-body method called the RCC theory which supplements the missing correlations in DF theory. \\

\begin{table}
\caption{The comparison of EDM enhancement factors of the ground state of Rb and Cs atoms. The references are given in the parenthesis.\\}

\begin{center} 
\begin{tabular}{|c|c|c|c|c|}
\hline
 Atom  &  Ground State &  \multicolumn{3}{|c|}{EDM enhancement factor, R} \\\cline{3-5}
       &               & \multicolumn{2}{|c|}{Our work} & Others \\ \cline{3-4}
       &               &  DF result & DF + Corr. & DF + Corr. \\
\hline
Rb&$4p^6\,5s\, (^2S_{1/2})$      & 19.6 & 24.6 &25.67 \cite{shukla} \\
\hline
Cs&$5p^6\,6s\, (^2S_{1/2})$&  94.2 & 125.9&114.2 \cite{hartley}\\
      &  &        &       &130.5 \cite{das}\\
      &  &        &       &120 \cite{sandars65} \\
\hline

\end{tabular}
\end{center}
\end{table}

In order to emphasize the contributions from the correlation effects to the total result, we have shown explicitly the individual contributions from the important RCC terms in the present calculation, in Table-2. It is evident from our results that, the electron correlation effects are quite important in the calculation of the EDM enhancement factors for the ground state of both Rb and Cs atoms.\\
\begin{table}
\caption{The individual contributions of the important RCC terms to the EDM enhancement factors of Cs and Rb atoms.\\}
\begin{center} 
\begin{tabular}{|c| c| c|}
\hline
{RCC term + c.c.} & Contribution (Cs) & Contribution (Rb)\\
\hline
      &  & \\
      $T1^{(1)^\dagger} \overline{D^{(0)}}$  & 5.33        &  1.05\\
      $S1^{(1)^\dagger} \overline{D^{(0)}}$  & 138.03         & 26.08\\
      $S2^{(1)^\dagger} \overline{D^{(0)}}$  & -7.15          & -1.02 \\
      $S1^{(1)^\dagger} \overline{D^{(0)}} S1^{(0)}$ & -2.28   & -0.39\\
      $S1^{(1)^\dagger} \overline{D^{(0)}} S2^{(0)}$ & -6.30   & -0.95\\
      $S2^{(1)^\dagger} \overline{D^{(0)}} S1^{(0)}$ & -0.57   & -0.07\\ 
      $S2^{(0)^\dagger} \overline{D^{(0)}} T1^{(1)}$ & 0.63    &  0.13\\
      $S2^{(0)^\dagger} D S2^{(1)}$ & 1.97       &  0.26\\
      $S2^{(1)^\dagger} \overline{D^{(0)}} S2^{(1)}$ & 0.32    &  0.04\\
      Rest of the terms & -1.13    & -0.14\\
      Normalization     & -2.91   & -0.42\\ 
\hline
      Total value       & 125.94  & 24.57\\
\hline
\end{tabular}
\end{center}
\end{table}
     
Our RCC calculation of the enhancement factors are in reasonable agreement with the previous calculations \cite{shukla, hartley, das, sandars65}. The calculations by Shukla et al. \cite{shukla} and Das \cite{das} are based on a hybrid method combining the salient features of the relativistic many-body perturbation theory (MBPT) and the multi-configuration Dirac-Fock (MCDF) method. Hartley et al. have used a variant of the relativistic MBPT \cite{hartley} to perform their calculation. Their treatment of pair correlation effects like the former two calculations is not as comprehensive as ours. Sandars' calculation is based on a simple relativistic one electron theory \cite{sandars65}. \\

As mentioned before, our calculated enhancement factors along with the results of the EDM experiments on atomic Rb and Cs when they come to fruition could provide the most sensitive limit for the electron EDM.\\ 

There are many non-standard models (non-SM) like the different super-symmetric models, left-right symmetric model etc., which predict the EDM of the electron to be almost in the same range, where as, the SM prediction is ten to twelve orders of magnitude less which is far off from the reach of current experimental facilities \cite{bernreuther}. If one can unambiguously observe the non-zero EDM of Rb and Cs atoms then one can combine it with our present calculation of the enhancement factors and choose the correct model of particle physics which predict the value of electron EDM in that range. It also constrains the parameter space of many models of CP violation. Thus, the work on electron EDM has the potential to probe a new physics beyond the SM.\\

\section{Conclusion}
We have calculated the electron EDM enhancement factors of two paramagnetic atoms such as rubidium and caesium in their ground states using a highly correlated all-order relativistic coupled-cluster theory. Our results can be used in combination with those of the proposed high precision EDM experiments to obtain the stringent limits for the electron EDM, there by shedding light on leptonic CP violation which hopefully will unveil a new arena of physics beyond the most celebrated model of particle physics till date, the standard model.\\


\section*{References}
\begin{thebibliography}{20}
\bibitem {ginges} Ginges J S M and Flambaum V V 2004 {\it Physics Reports} {\bf 397} 63\\
\bibitem{maxim} Maxim Pospelov \& Adam Ritz 2005  {\it Preprint} hep-ph/0504231\\
\bibitem {sandars} Sandars P G H  and Lipworth E 1964 {\it Phys. Rev. Lett.} {\bf 13} 718\\
\bibitem {sandars2} Sandars P G H 1968 {\it J. Phys.} B {\bf 1} 511\\
\bibitem {kazarian} Kazarian A M, Kuzmin S V and Shaposhnikov M E 1991 {\it Preprint: } CERN Report {\bf CERN-TH.6132/91}\\
\bibitem {branco} Branco G C, Luis Lavoura and Silva J P 1999 {\it CP Violation} (Oxford: Clarendon Press) \\
\bibitem {schiff} Schiff L I 1963 {\it Phys. Rev.} {\bf 132} 2194\\
\bibitem {salpeter} Salpeter E E 1958 {\it Phys. Rev.} {\bf 112} 1642\\
\bibitem {regan} Regan B C and Commins E D 2002 {\it Phys. Rev. Lett.} {\bf 88} 071805\\
\bibitem{sudha} Murthy S A, Krause D, Li Z L and Hunter L R 1989 {\it Phys. Rev. Lett.} {\bf 63} 965\\
\bibitem {weiss} Weiss D S, Fang F and Chen J 2003 {\it Bull. Am. Phys. Soc.} {\bf APR03} J1.008\\
\bibitem{heinzen} Heinzen D J {\it Private communication}\\
\bibitem {lindgren} Lindgren L and Morrison J 1986 {\it Atomic Many-Body Theory} (Berlin: Springer-Verlag)\\
\bibitem {mukherjee} Mukherjee D and Pal S 1989 {\it Adv. Quantum Chem.} {\bf 20} 561\\
\bibitem {kaldor} Kaldor U 1987 {\it J. Chem. Phys.} {\bf 87} 4678\\
\bibitem {rajat} Chaudhuri R K, Panda P K and Das B P 1999 {\it Phys. Rev.} A {\bf 59} 1187\\
\bibitem{shukla} Alok Shukla, Bhanu Pratap Das, and Andriessen J 1994 {\it Phys. Rev.} A {\bf 50} 1155\\
\bibitem{hartley} Hartley A C, Eva Lindorth and Pendrill A M M 1990 {\it J. Phys.} B {\bf 23} 3417\\ 
\bibitem{das} Bhanu Pratap Das  1989 {\it Aspects of many-body effects in molecules and extended systems} {\it Lecture notes in Chemistry} {\bf 50} 411 (Berlin Heidelberg: Springer-Verlag)\\
\bibitem{sandars65} Sandars P G H 1965 {\it Phys.Lett.} {\bf 14} 194\\ 
\bibitem{bernreuther} Werner Bernreuther and Mahiko Suzuki 1991 {\it Reviews of Modern Physics} {\bf 63} 313\\
\end{thebibliography}
\end{document}